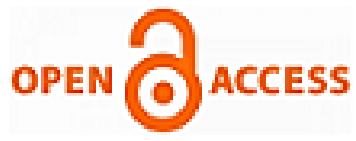

# Travel package booking application with API BOT

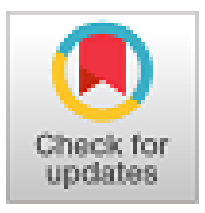

**K Sai Karthik, CH Naveen Aaditya, Ravi Kiran, Swarnalatha P**

***Abstract:** These days we are witnessing many mobile applications based on the recommended systems, which have become a great technology which is been used by the various mobile applications according to the situation. Recommendation provided by the mobile application is a key element for the person who is traveling to several places. For any tourist information application contextual information is much needed to guide the user on his interests this can be achieved by the Context-aware computing. Which provides the user most interactive system with the suggestions provided by it based on the input from the user in a certain location, here context includes the user's mental, social, physical environments. To achieve this contextual information, we will design and implement the context-aware user interface based on the user for which we have to study the user and design a rich user interface. The final outcome for which users have the satisfaction when using context-aware functionality will be much better than non-context-aware application.*

## I. INTRODUCTION

A bot also called as web robot is a software application. It runs automated tasks when connected to the internet. It analyses and gets information from web server faster than the human speed. These bots are used to communicate with the users through instant messaging and also through Facebook bots, telegram bots and Twitter bots. The bots can handle many things such as providing information about weather forecast, cricket scores, ZIP code information and many other data. This is very useful to the people to get the information. Companies benefit a lot from bots. Bots are helpful because they allow people to communicate with them without the

Person involved in it. Customers interaction has increased after the development of chatbots as they communicate faster than the humans. Google assistant is the most used chat bot now a days. Google assistant is benefiting people is their day to day life.

* Correspondence Author

**K Sai Karthik* ,** School Of Computer Science And Engineering, Vellore Institute Of Technology,Vellore, Tamil Nadu, India.

**Ch Naveen Adithya**, School Of Computer Science And Engineering, Vellore Institute Of Technology,Vellore, Tamil Nadu, India.

**Ravi Kiran**, School Of Computer Science And Engineering, Vellore Institute Of Technology,Vellore, Tamil Nadu, India.

**Dr. Swarnalatha P**, School Of Computer Science And Engineering, Vellore Institute Of Technology,Vellore, Tamil Nadu, India.

In this project bot is used to know about the current available packages, to book ticket to any available package, and also responds to the problems or issues by providing the necessary solutions. Bots are also used in trading business like eBay and many other trading websites.

After Facebook, Twitter allowing the chat bots to chat in their platform there has been a tremendous growth in the use of chat bots. We can also download chat bot from google play for android users and Apple App store for iOS users. By using chat bots, the money spent will be very less compared to the money spent without using chat bots. So many companies are using bots frequently these days. Social networking bots is also one type of bot which is used for repetitive set of instructions. Social networking bots are designed in such a way that it responds to the question asked by the customer similar to the way the customer speaks. Bot is also called as intelligent agent.

There are many types of bots like Chat bot, Internet bot, Social bot, Spam bot and many types of bots. Bots are generally used to do repetitive tasks as human cannot do the repetitive tasks all the time. Some bots are not useful as there is no security involved in it. Some people also cracks the users accounts and make benefits out of it which is very harmful. Bots can scan all the information available on the internet in all web pages. Bots which are harmful comes under bad bots. Bots which are used to hack the accounts which leads to cybercrime is one of the example of bad bots. Bots should be managed properly as they may lead to many problems if they are not maintained properly. Too much access of bots can also slow down the response from the bot as the traffic is higher. The bot manager is the software that manages the bots. Bot managers should block the bad bots which are very harmful. Many people use google bots as they are very secured and there is a very good organic traffic in it. Bot manager analyses bot behaviour, add good bots to white lists, identify bot reputation and many such things. Creative tasks are not done by bot. only non-creative tasks are done by the bot. Creative tasks are done by the humans. Bots are employed on a large scale. Bot is the software that became easier to implement. It makes user feel a great experience while interacting with it.

## II. PROCEDURE FOR PAPER SUBMISSION

1) The main objective of this research is to develop a mobile travel guide application with added functions to an existing application.

- The application itself finds out your current location.
- Application suggests the person desired places and events.

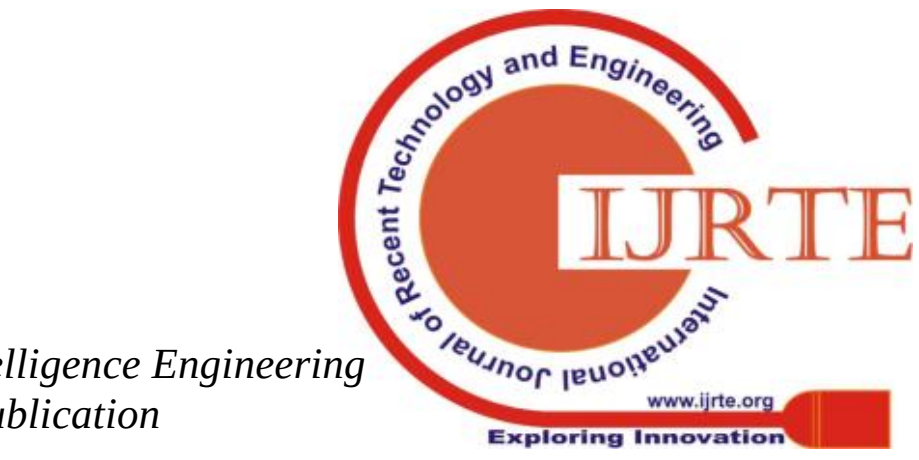

- Also provides the Reviews on that particular place.

2) Collect the information regarding for the specific place.
- Through some questionnaires.
- Posting some interview kind of questions.

3) Providing an option through our application to post reviews based on his personal interests.

4) Online ticketing and online shop sale services allow users to buy tickets and souvenirs anywhere with an access to a network.
- Good user interface for ticket booking.
- Providing more convenient ways for payment.
- Gives an option to print it as a hard copy.
- Also provides a QR code for that ticket.

5) The application provides some gestures to make it more convenient for the users.

- Long press anywhere on the item detail page bookmark interaction method.
- One-finger-hold-pinch is used to review the page or place.
- Tilt the phone for getting back to the previous pages

## III. RELATED WORK

From the journal N. Davies, K. Cheverst, A. Dix in Computing Department, Infolab21 Lancaster University Hesse in Technische Universität Ilmenau, Ilmenau, Germany [16]we understand that the growth of context-aware computing has led to the development of location-dependent applications, such as mobile tour guide systems. Although providing regularly updated information about users' current location, as well as displaying detailed information about specific features linked to their position, relatively limited work has been done investigating the impact of context-aware information on system usability.

By the paper Unsupervised feature selection using weighted principal components from Kim, S. B. and Rattakorn, P., K. Laakso, O. Gjesdal and J.R. Sulebak we came to know that availability of immense amount of digital information makes it difficult for users to find their desired content(e.g., movies, books) in a reasonable time. Contemporary society today has entered the era of Big Data.Because of the increasing amount of available digital information. Hence, such a problem of excessive information prompted a phenomenon called information overload. Recommender systems solve this problem by searching through a dynamically generated large volume of information to provide content and services based on the user's taste and preferences.

Analysing the paper Location- and Time-Based Information Delivery in Tourism by A. Hinze and A. Voisard we came to know that Advanced tourist information system provide targeted and essential data that is semantically-rich to mobile users, based on the user's choices and travel history. Some systems also recommend sights that match the user's context. System for mobile tourist information, considers the personal background of a traveler both for selecting the information that is delivered to the user and for recommended routes.User context also includes their interest, location, and means of travel. A user's interests are captured by the questionnaire. The systems also take selected aspects of the semantic context of sights into account (e.g: their location, their type, and similarity to other sight.

From the paper Context aware recommendation systems, incorporating relevant contextual information when generating or providing recommendations improve the accuracy of prediction and efficiency of the recommendation. A type of recommender system that utilizes contextual information to adopt its recommendations to users' contextual situations is known as context-aware. A context-aware recommendation system labels each action of the user with an appropriate context and effectively adopts the system output to the user in that given context.

## IV. USERS CLASSES AND CHARACTERSTICS

### A. System administrator

Adding/editing of the products in database is done by administrator. He can also change user interface for the website. He can also make changes in database.

### B. Customer

Customer can book packages in website using website interface and can get the products ordered.

### C. API BOT

API BOT has been used to website and it linked to a bot named travel in telegram so that user can solve any query by using this.

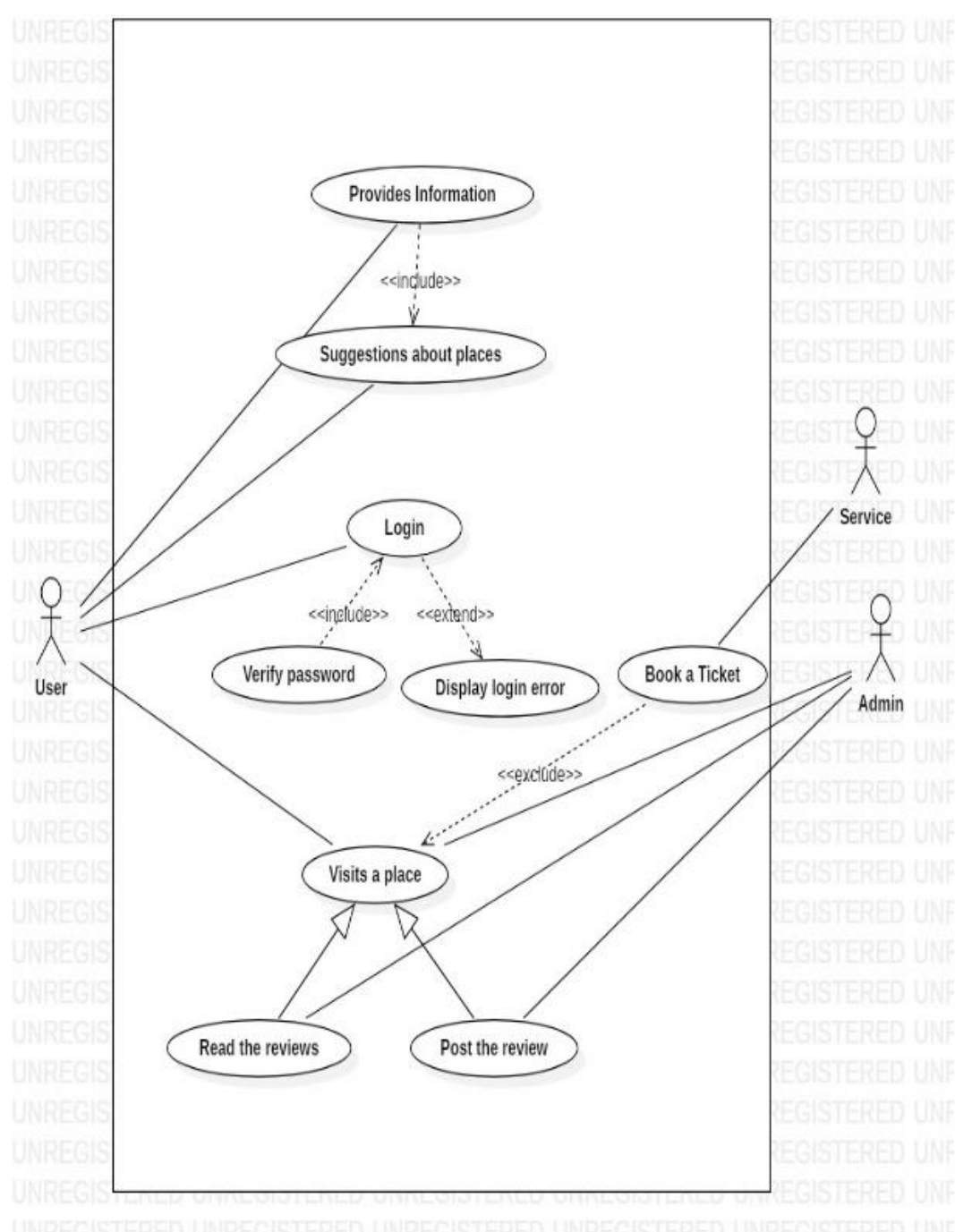

**Fig 3.1 Represents the use case diagram of the project.**

## V. APPLICATION HARDWARE AND SOFTWARE OVERVIEW

### A. Operating Environment

- This website works on any web browser using any electronic gadgets.
- For proper functioning of system it usesDDS.
- Database of this website will run using SQL.
- It is developed on a platform of HTML, CSS, PHP, Bootstrap, Javascript.

### B. Hard ware interface

To use this website, the system must be connected to internet through WAN-LAN or mobile data.

### C. Software interface

It requires scripting languages like VBSCRIPT, PHP etc., as the system is working with a server.
To make any changes in database and for storing the data during transaction requires MYSQL.

### D. Communication interface

Domain name space is needed for naming on the internet. It also requires Web Applications like google chrome, Mozilla Firefox, opera mini, Internet Explorer etc., for proper interaction with the system.

### E. Different aspects of Website

- Registration: Customers who want to buy tourist packages can initially sign up through registration and later book their tickets through their account. Admin can add the packages available so that the customers can see that by signing up. Packages tours who wants to provide booking the packages in travel guide application can register through website with valid proof such as Aadhaar card.
- Login: Registered users/vendors/delivery boy can login with their login credentials.
- Changes to cart: customers can add or delete the packages from the cart using their login credentials.
- Status of booking: Customers can find their status using the website.

### F. Performance of website

- Related Searching*:* This website can process on search queries and inform the customers with relevant information.
- Distributed system: This website can be used on any relevant gadget and on basis of server strength only a limited user can use this website.
- End-user efficiency: Users of this website can send their queries and feedback for updating of website through mail.

### G. Usability

- Documentation: The website is completely documented so that specific users can easily navigate the products.
- Aesthetics: The website is clean and has stylish look. As website designed in such a way that loading time is minimized for fast browsing, so that many products can be navigated in less time.

### H. Reliability

- Recoverability: The website has a backup hard drive, so that in case of failure of primary hard drive there will be no loss of data.
- Predictability: Maintenance of server is done during hours of minimal traffic.

### I. Supportability

- Compatibility: This website is compatible with proper functioning web browser.
- Maintainability: Administrators can add/edit products on their convenience. Admins maintains the entire website. Administrators validates the products posted by vendors in website for proper functioning of website.

## VI. FUNCTIONALITY OF THEWEBSITE

A registered /consulted hotel manager of our website can login into website using his login credentials. After logging in he gets the booking details to be confirmed. He/she can also add products of his/her interest in website using their credentials which will be appearing in the Booking page used by customers for surfing of packages.

On booking a product by customer, it notifies the vendor who is closer to customer through email and our website.

A registered customer can login into website through their login credentials. After logging in, Customer surfs for the required products and add them to his/her cart. After adding products in cart, he can check out for the bill and place the order by adding his phone number and address details. The placed booking directs to the confirmation.

## VII. IMPLEMENTATION

### A. Login Panel

Login page is where the user enters his user name and password to login to the website.

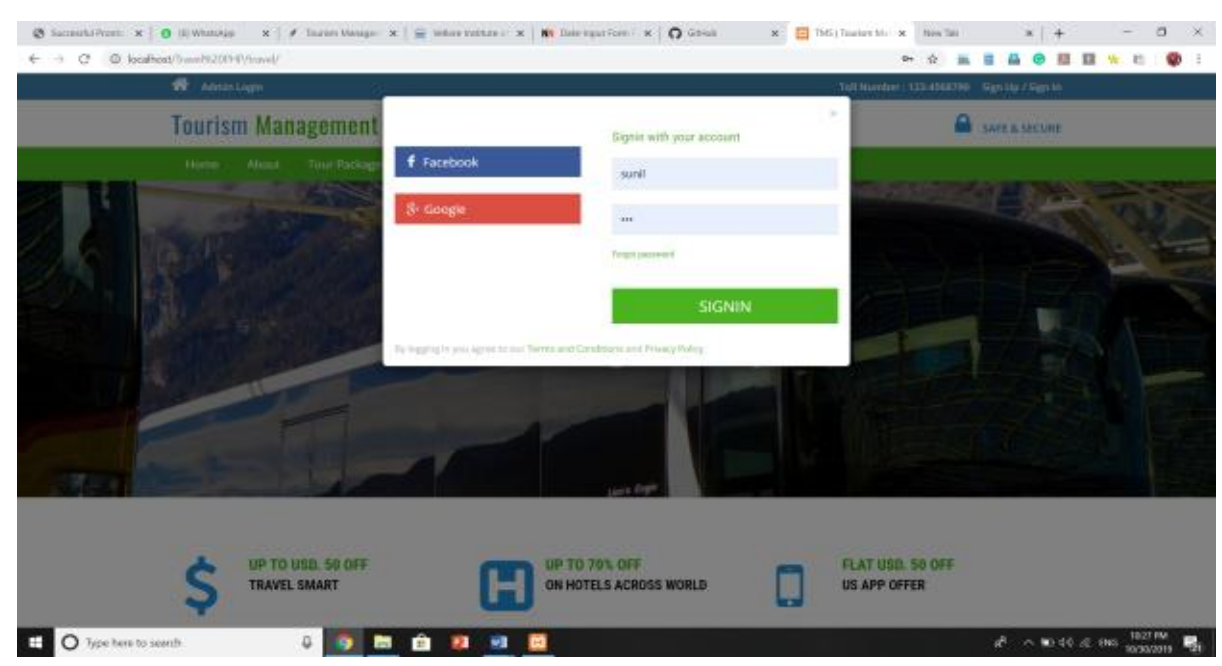

**Fig 6.1 Represents login Panel.**

### B. Package list

Package list is the one which the customers can select their packages

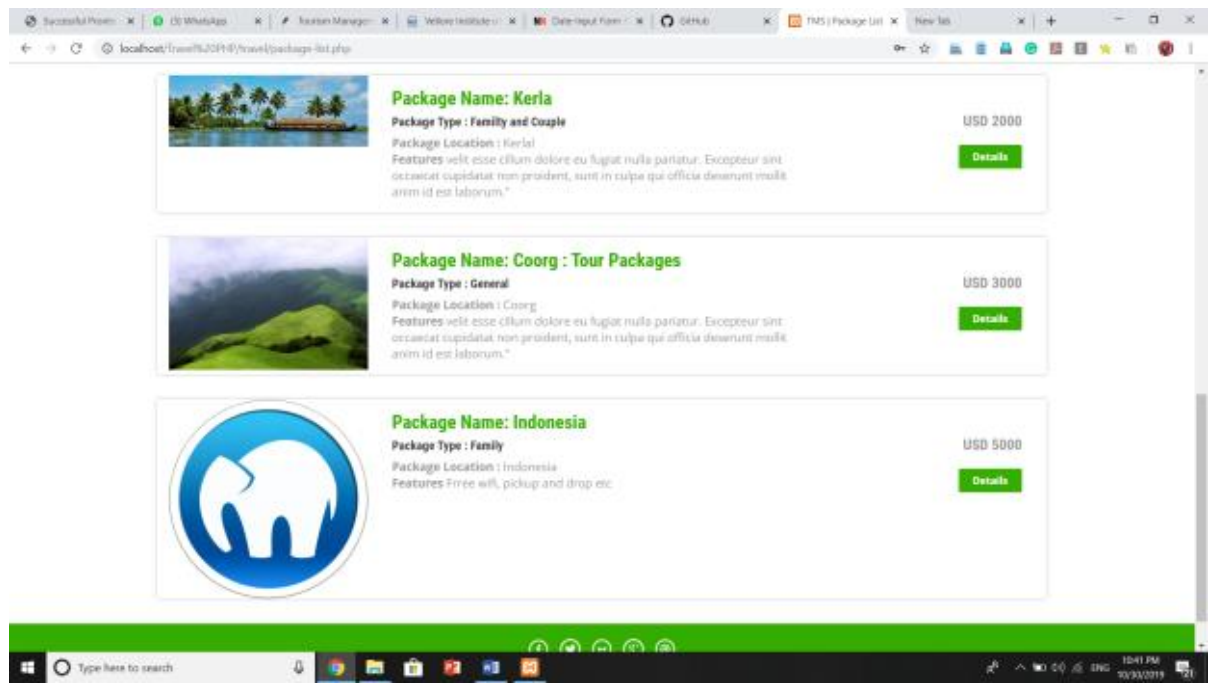

FIG 6.2 Represents package list

### C. Booking details

Customers can book their tickets from this page. Customers can view the amount to be paid.

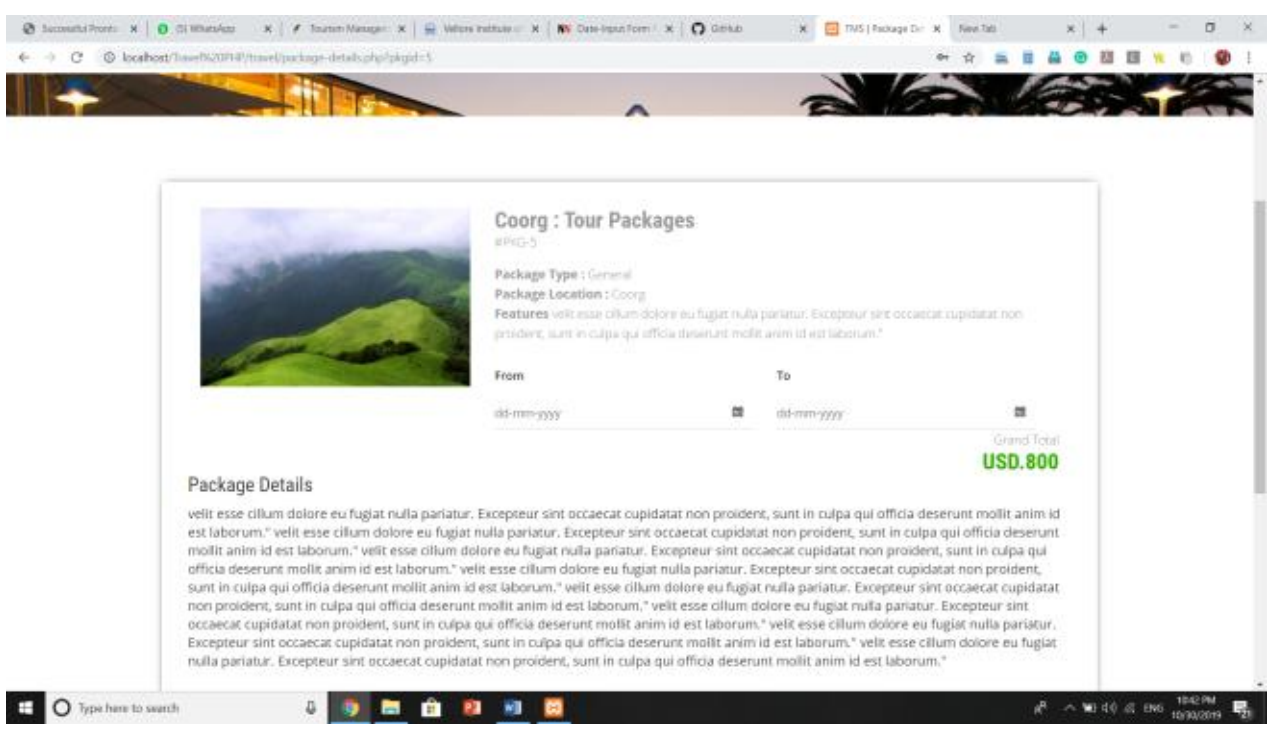

FIG 6.3 Represents booking details

### D. Feedback page

After visiting to that place user can give his feedback by this page about the experience of the tour.

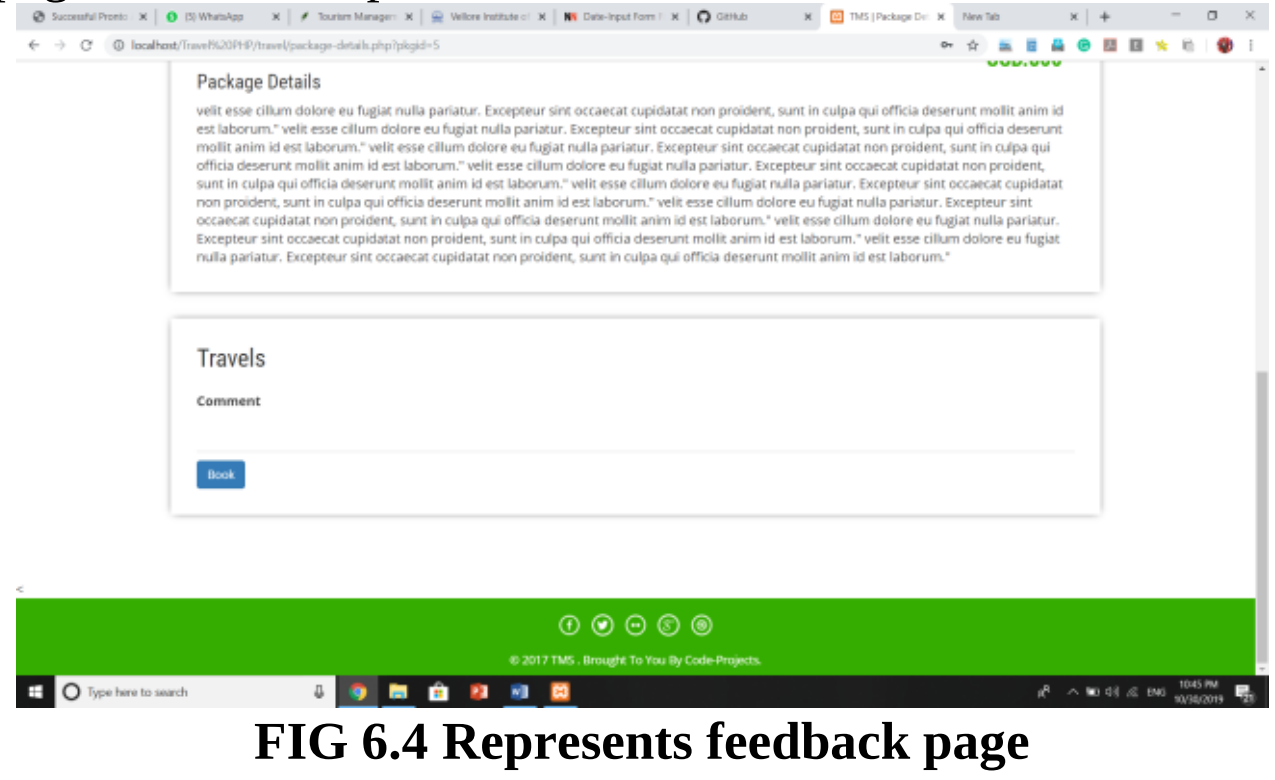

FIG 6.4 Represents feedback page

### E. Chat bot page

It will give guidelines to the customers,helps to use our website efficiently,gives information about which place has more ratings,books tickect in less time and takes upto the payment process.Mainly it develops interaction between user and system.

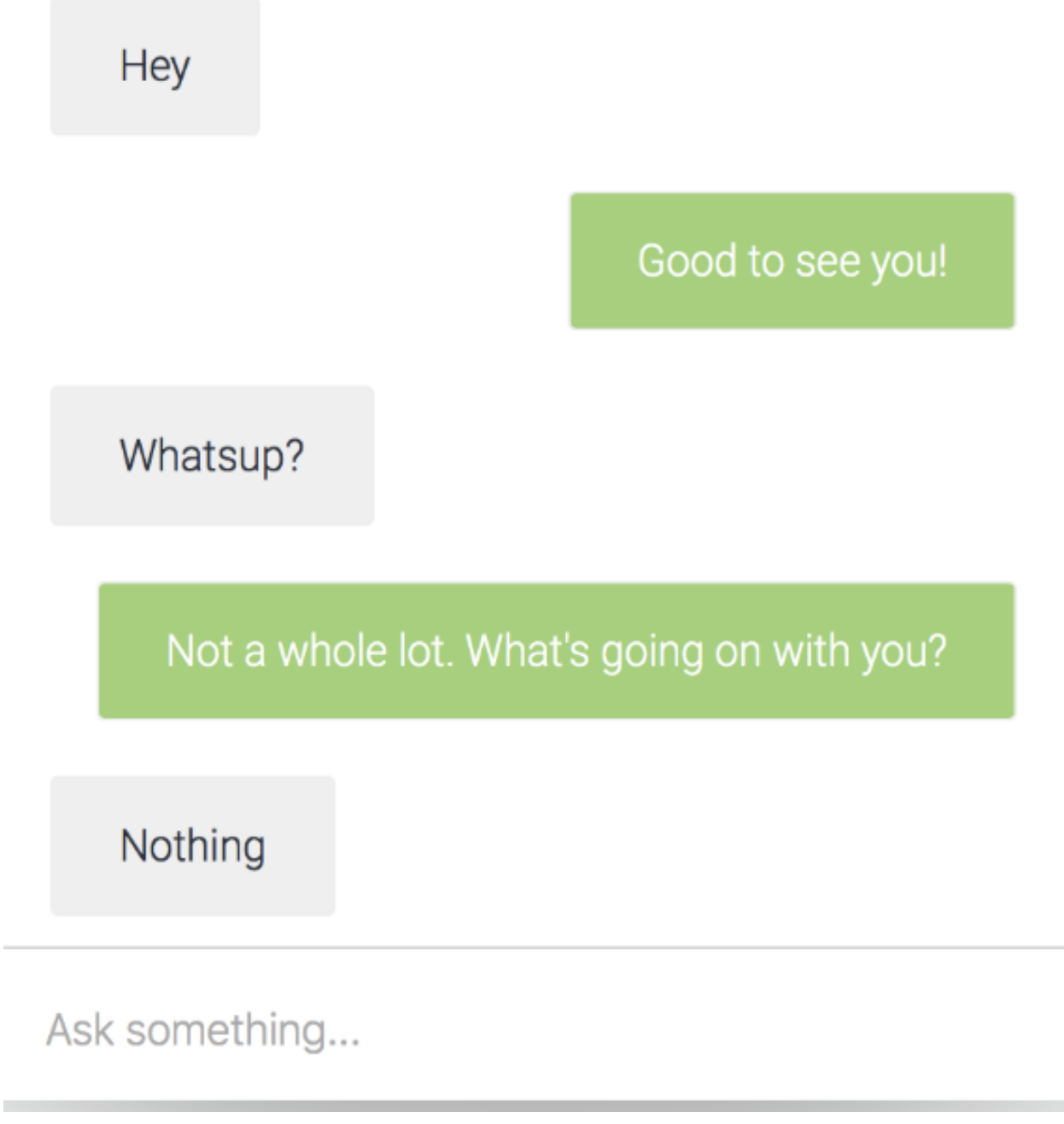

FIG 6.5 Represents Bot

## VIII. PERFORMANCE ANALYSIS

In existing system there is no bot present. In our proposed system, bot is present which is used to interact with the customers. Context aware functionality is present in our proposed system. Application suggests the users desired places and events. Users interests are identified by posting some interview kind of questions. Application itself finds the current location by itself and responds accordingly. Bot is also linked with the telegram where you can communicate there itself without signing into the website. The response by the bot in telegram and website is same. In telegram we also provided the response in such a way that the pictures are displayed indicating the response of the question.

## IX. TELEGRAM BOT

As we linked our bot with a bot in telegram so that we can access our site by using telegram not by using website every time. So by using telegram bot we can know the available packages so that we can plan for it before signing in into the website. Whatever the bot responds in the website same response is given by the bot in the telegram also.

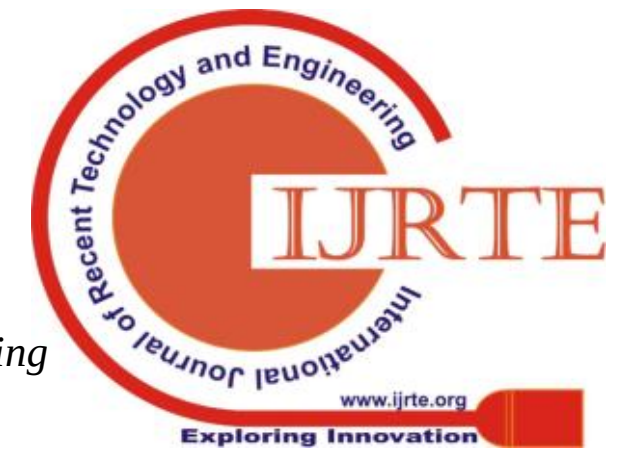

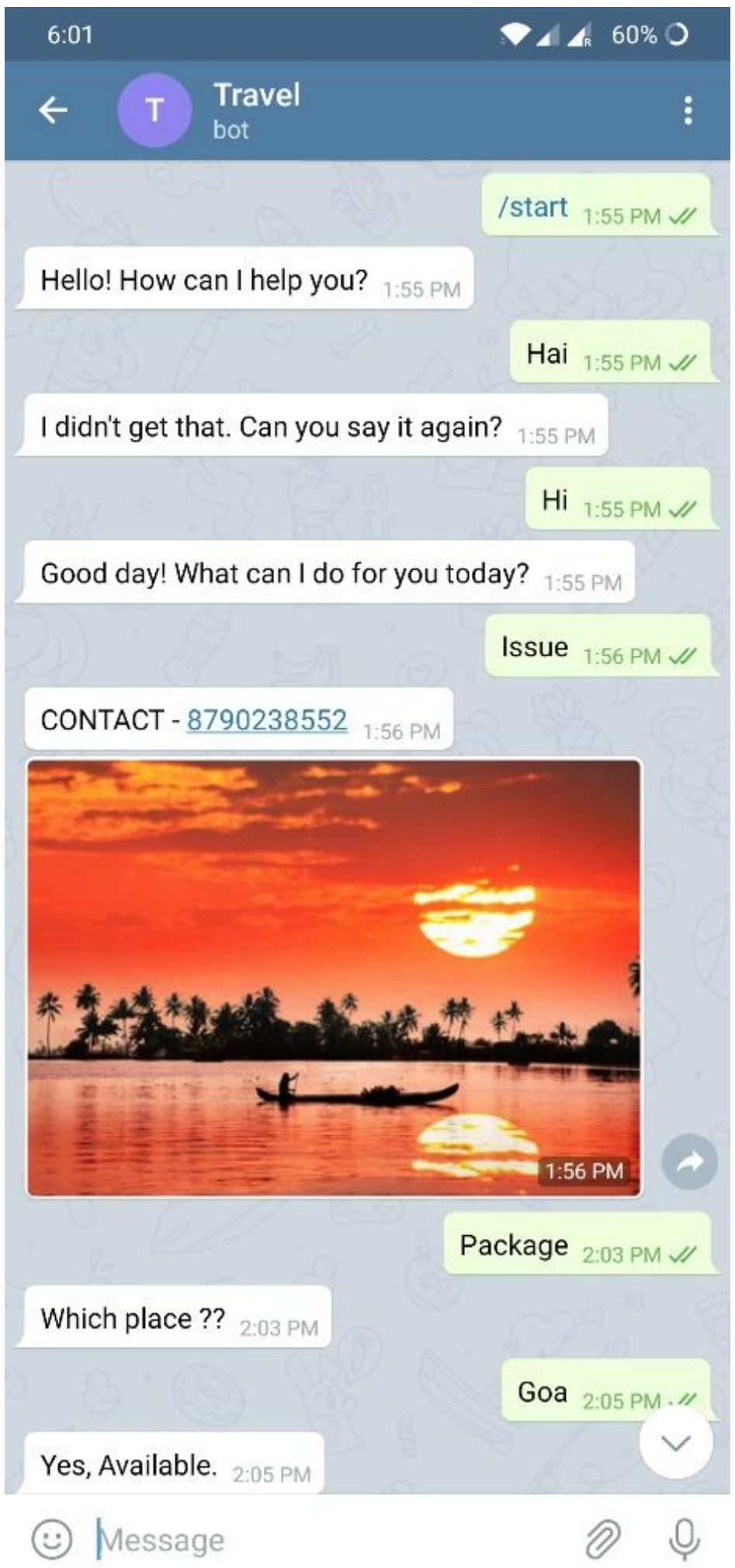

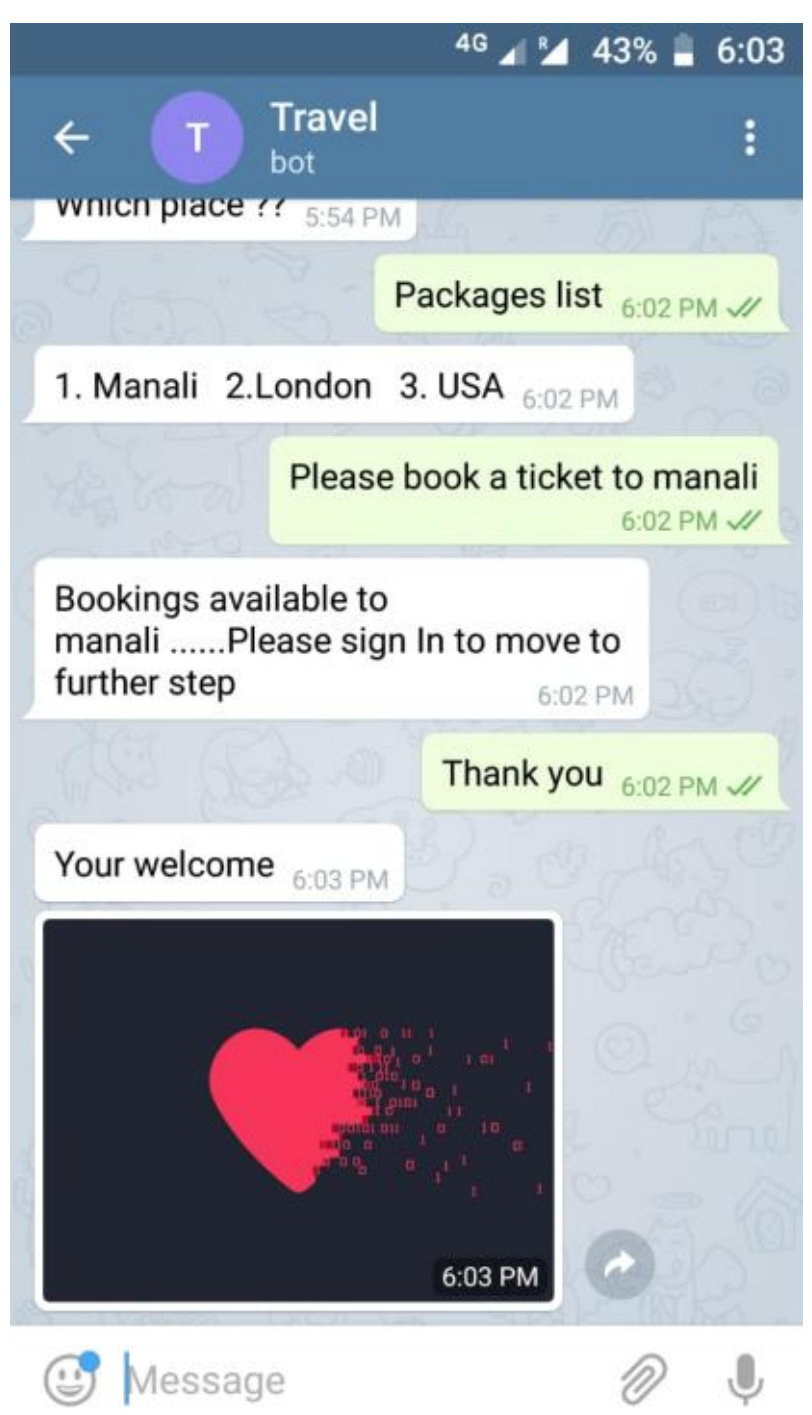

**FIG 10.1 and 10.2 Represents Telegram bot chat**

## X. CONCLUSION

In this project, we have developed a mobile travel guide application using context aware computing which provides the user most interactive system with the suggestions provided by it based on the input from the user in a certain location, here context includes the user's mental, social, physical environments. We have designed and implemented the context aware user interface based on the user by designing good user interface. We have developed in such a way that the user will have satisfaction when using context-aware functionality which is much better than non-context-aware application. We have provided the option through our application to post reviews based on his personal interests so that the new users can see the reviews given by the other people and plan accordingly. We have implemented the context aware computing by posting some interview kind of questions to the user so that the application suggests the user desired places and events by finding the current location. We have also provided the good interface for ticket booking and more convenient ways for payment. We also include the voice recognition where the user can ask questions regarding the application and also responds the answer related to it.

## ACKNOWLEDGEMENT

This project provides us an opportunity to have a greater understanding of the subject and explore beyond the book. On this occasion, we would like to express our heartfelt gratitude to our faculty, for his guidance and encouragement throughout this project.

## AUTHORS PROFILE

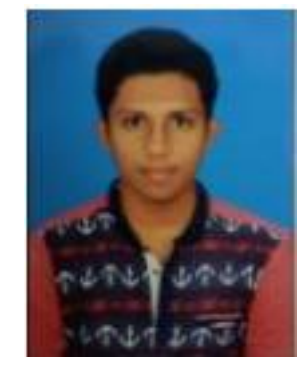

**K Sai Karthik** , He is pursuing under graduation in the stream of Computer science and Engineering at Vellore Institute of Technology, Tamil Nadu, India.

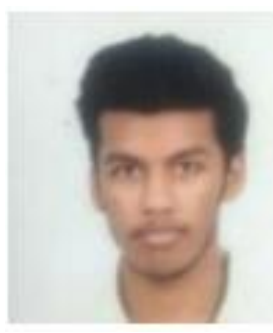

**CH Naveen Aaditya**, He is pursuing under graduation in the stream of Computer science and Engineering at Vellore Institute of Technology, Tamil Nadu, India.

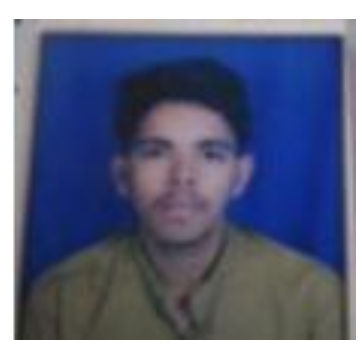

**Ravi Kiran** ,He is pursuing under graduation in the stream of Computer science and Engineering at Vellore Institute of Technology, Tamil Nadu, India.

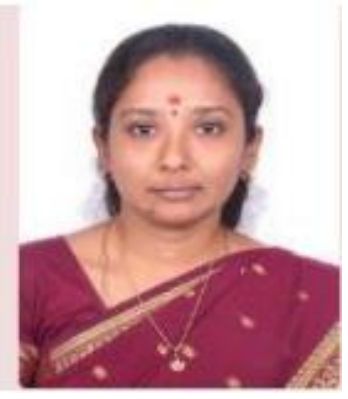

**Dr. Swarnalatha P** is currently working as an Associate Professor in the Department of Computer Science and Engineering, VIT Vellore India. He completed his Ph.D. from the Department of Computer Science & Information Systems. Her work on Image Processing for medical and satellite images bought her a funded project on “Development of approaches for geometric, radiometric and atmospheric correction of remote sensing data projects for multi-date data analysis” by Space Application Centre-Ahmedabad at VIT University, Vellore and consultancy work for CMC, Vellore. She is a Member of many Professional Societies such as Life Member of Computer society of India (CSI) Professional Member of Association of Computing Machinery (ACM), Senior Member of IEEE and Member of ACEEE.

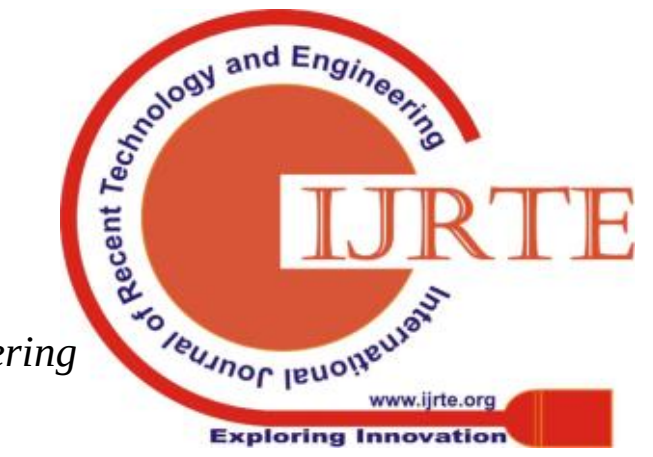